\documentclass[twocolumn,prl,showpacs,superscriptaddress]{revtex4}

\usepackage[dvips]{graphicx}
\usepackage{amssymb,amsfonts,amsmath}
\usepackage{graphicx}  % needed for figures
\usepackage{dcolumn}   % needed for some tables
\usepackage{bm}        % for math
\usepackage{amssymb}   % for math

\usepackage{color,soul} % per evidenziare
\begin{document}

\title{Single molecule study of the DNA denaturation phase transition in the force-torsion
space}

\author{D. Salerno}
\author{A. Tempestini}
\author{I. Mai}
\author{D. Brogioli}
\author{R. Ziano}
\author{V. Cassina}
\author{F. Mantegazza}
\affiliation{Dipartimento di Medicina Sperimentale, Universit\`a degli Studi di Milano - Bicocca, Via Cadore 48, Monza (MB) 20900, Italy.}

\date{\today}

\begin{abstract}
We use the ``magnetic tweezers'' technique to reveal the structural
transitions that DNA undergoes in the force-torsion space.  In
particular, we focus on regions corresponding to negative supercoiling. These regions are
characterized by the formation of so-called denaturation bubbles,
which have an essential role in the replication and transcription of DNA.
We experimentally map the region of the force-torsion space where the
denaturation takes place. We observe that large fluctuations in
DNA extension occur at one of the boundaries of this region, i.e., when
the formation of denaturation bubbles and of plectonemes
are competing. To describe the experiments, we introduce a suitable extension of the classical model.
The  model correctly describes the position of the denaturation regions, the
transition boundaries, and the measured values of the DNA extension
fluctuations.
\end{abstract}


\maketitle

The nanomechanics of DNA play an important role
at the biological and biochemical levels \cite{book_understanding_DNA}.
 Thus, understanding the transcription and duplication phenomena is a relevant open topic to which a quantitative comprehension of DNA mechanical characteristics is fundamental. In particular, because any transcription or duplication process implies the local and temporary separation of the two DNA strands (i.e., DNA breathing \cite{lee2010,theodora2012} or denaturation bubbles \cite{erp2005,metzler2009,Nisoli2010}), understanding denaturation represents the first building block towards the theoretical comprehension of DNA metabolism.
A well-known and promising technique for studying nanomechanical properties
is the magnetic tweezers (MT), which allows one to impose a
stretching force and a torsion to a single DNA molecule while also monitoring the simultaneous extension of the
same molecule \cite{neuman2008, strick1998}.
The versatility of the MT technique has been exploited to
investigate DNA nanomechanics in the presence of proteins, enzymes, ligands, and
drugs \cite{koster2005,koster2007,nollmann2007, salerno2010, lipfert2010}, and phenomenologically analyzed \cite{allemand1998}.
The initial pioneering MT studies focused on the
topology of DNA molecules and showed that torsion can
produce a so-called ``plectoneme'', which reduces DNA extension
\cite{smith1992,strick1996}. For modeling plectoneme formation, the
DNA can be simply described as an elastic rod \cite{strick2003}.
The experiments showed that the plectonemes disappear when the force becomes sufficiently
high and the direction of the torsion is toward the unwinding of the DNA
double helix \cite{strick1998}.
This chiral effect, which goes beyond the elastic rod
model, has been explained in terms of denaturation of the double
helix \cite{allemand1997}.

In this work we use the asymmetry between the DNA extension under positive and negative torsion as a hallmark of denaturation.
For the first time, we systematically evaluate the occurrence of mechanical denaturation in the force-torsion space. We find that large temporal fluctuations of extension arise at one of the boundaries of the denaturation region.
Finally, we interpret the experimental data with a simple mechanical
model obtained by considering a denaturation term to the classical energy \cite{strick2003} used
to describe  the buckling transition.

Several MT apparatuses have already been reported in the literature and, in our
set up \cite{salerno2010}, we generally follow the most classically proposed schemes
\cite{strick1998, gosse2002}. The technique is based on
the following procedure: one end of the DNA is connected
by standard biochemical techniques \cite{strick1998} to a commercial
micron-sized superparamagnetic bead and the other DNA end is fixed to
the inner wall of a squared capillary tube \cite{lipfert2009, teVelthuis2010}.
Forces or torsions are then applied to the bead
through a field generated by external permanent magnets,
whose position and rotation can be
controlled \cite{klaue2009, mosconi2009}. The movement of the magnetic bead is
transferred to the DNA and thus, indirectly, forces or torsions are applied to the molecule \cite{dessinges2002}.
The DNA extension $L_e$ is measured by
considering the diffraction images generated at different heights of
the bead, which is illuminated by a LED light
\cite{gosse2002}.
The force $F$ exerted by the magnetic field on the bead and, as consequence exerted on the DNA, is obtained as in ref. \cite{lipfert2009}.

\begin{figure}
\includegraphics[width=.4\textwidth]{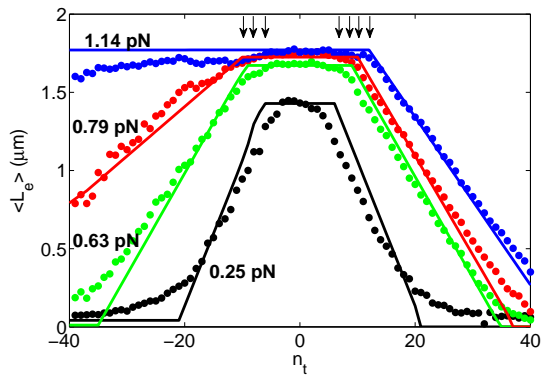}
\caption{Average DNA extension $\langle L_e\rangle$ as a
  function of the number of imposed turns $n_t$.
 Data (\emph{dots}) and theoretical curves (\emph{lines}) are obtained for different values
  of the applied force $F$, as indicated in the
  figure. The short vertical arrows point to the
  buckling transitions $n_b$.}
\label{chapeau}
\end{figure}

Figure~\ref{chapeau}   shows the average value of the extension $\langle L_e\rangle$ of a DNA ($\sim 6000$ base pairs)
molecule as a function of the number of imposed turns $n_t$ at different
values of the applied stretching force ($F$~=~0.25~pN, 0.63~pN, 0.79~pN, and 1.14~pN).
The data are the result of an averaging
procedure on several values of $L_e(t)$ taken as a function of time at
a frequency rate of 60~Hz during a time interval of several seconds.
These results are typical in the literature of MT
\cite{strick2003} and are qualitatively described as
follows.
When increasing the absolute value of $n_t$,
the torsion is first absorbed by elastic twist deformations and the
DNA extension remains approximately constant.
Above the so called $n_b$ buckling transition, dependent on $F$
and indicated by the vertical arrows in Fig.~\ref{chapeau},  the creation of plectonemes
induces a progressive linear decrease in the DNA extension.
At low forces ($F$~$<$~0.6~pN), when increasing $n_t$, the data show a symmetric trend at positive and negative torsion.
At high force values ($F$~$>$~1~pN), the
situation is different: at negative imposed turns,
the $L_e$ vs $n_t$ curves are no longer symmetric, and
the DNA extension is approximately constant due to the formation of
so-called denaturation bubbles \cite{allemand1997}.

\begin{figure}
\includegraphics[width=.4\textwidth]{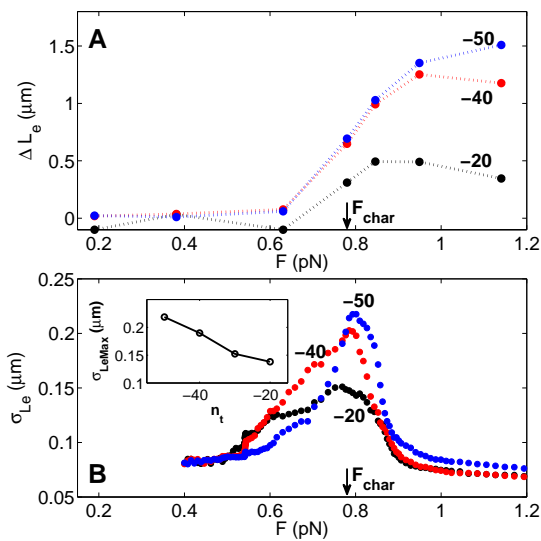}
\caption{Asymmetry $\Delta L_e(n_t)=\langle L_e(-n_t)\rangle-\langle L_e(n_t)\rangle$  (\emph{Panel A}) and
standard deviation  of the DNA extension $\sigma_ {L_e}$ (\emph{Panel B}).
  measured as a function of the applied
  force $F$. Data obtained for different values of  the imposed turns
  $n_t$ (indicated in the figures).  Inset: measured maximum values $\sigma_{L_eMax}$ obtained from the
  $\sigma_{Le}$ vs $F$ curves shown in the main
  figure.}\label{versus_F}
\end{figure}

The asymmetric behavior of the $\langle L_e\rangle$ vs $n_t$
is quantified by the expression
$\Delta L_e(n_t)=\langle L_e(-n_t)\rangle-\langle L_e(n_t)\rangle$.
The $\Delta L_e(n_t)$ values are shown in Fig.~\ref{versus_F}A
as a function of $F$.
 We observe a transition between the plectonemic behavior at low forces and the formation of denaturation bubbles at high forces. This transition is highlighted by an increasing asymmetry and occurs around a characteristic force $F_{char}$~$\approx$~0.78~pN, which does not depend
 on $n_t$.

At the intermediate forces (0.6~pN~$<$~$F$~$<$~0.9~pN) of Fig.~\ref{chapeau},
we have also observed that the average extension $\langle L_e\rangle$ at negative values of $n_t$ ($n_t$~$<$~-20)
appears noisier than the data for a positive $n_t$.
We have quantified the fluctuations by
calculating the standard deviation of the DNA extension $\sigma_{Le}
=\sqrt{\langle L_e^2\rangle - \langle L_e\rangle^2} $. We have also
checked that $\sigma_{Le}$ was not dependent on the total time of the
average and that such time was much longer (more than 100
times) than the characteristic time of the correlation
function of the data.
The values of $\sigma_{Le}$  as a function of $F$ for various values of the
number of imposed turns $n_t$ are shown in  Fig.~\ref{versus_F}B.
An increase in the extension fluctuations $\sigma_{Le}$ occurs
in a narrow range of applied force and presents a
maximum value $\sigma_{L_eMax}$ around  the characteristic force $F_{char}$:
the fluctuations are associated with the denaturation transition.
The value $\sigma_{L_eMax}$ is an increasing function of
$|n_t|$. This is confirmed in the inset of Fig.~\ref{versus_F}B, where the
obtained values of $\sigma_{L_eMax}$ are plotted for different
values of $n_t$ and show a linear behavior.

\begin{figure}
\includegraphics[width=.4\textwidth]{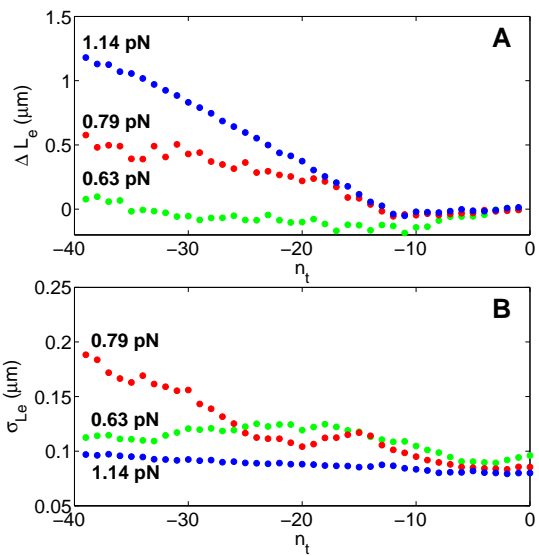}
\caption{Asymmetry $\Delta L_e(n_t)=\langle L_e(-n_t)\rangle-\langle L_e(n_t)\rangle$  (\emph{Panel A}) and
standard deviation  of the DNA extension $\sigma_ {L_e}$ (\emph{Panel B}) measured as  a function of the  imposed turns
  $n_t$.
  Data obtained for different values of the applied
  force $F$   (indicated in the figures).}
\label{versus_nt}
\end{figure}

We present the same quantities, $\Delta L_e$ and $\sigma_{Le}$, as a function of $n_t$ for different values of force $F$ (see Fig.~\ref{versus_nt}).
 We observe that the $\Delta L_e$ increase with negative supercoiling starting at $n_t$~=~-10 (panel A). This transition is not accompanied by an increase in $\sigma_{Le}$ (panel B). Instead, the strongest fluctuation takes place at the $F_{char}$, as shown previously.

\begin{figure}
\includegraphics[width=.4\textwidth]{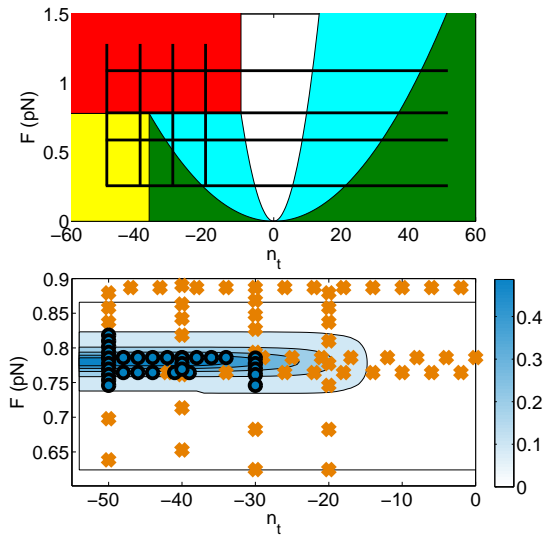}
\caption{\emph{Panel A}: Calculated phase diagram of DNA  structure in the ($n_t$,$F$)  space.
\emph{White}: DNA compatible with
the WLC model.  \emph{Blue}: plectonemic regions. \emph{Green}: zero DNA extension regions.
\emph{Red}: denaturation bubbles region. \emph{Yellow}:  coexistence region of denaturation bubbles and plectonemes at zero DNA extension.
 The vertical and horizontal lines correspond to the measurements presented at constant $F$ (in Fig.~\ref{chapeau} and Fig.~\ref{versus_nt}) and at constant $n_t$ (in Fig.~\ref{versus_F}), respectively.
\emph{Panel B} (enlargement of panel A): contour plot of the calculated (\emph{level
  lines})  $\sigma_{Le}$.
  Crosses (circles) indicate measured  values of $\sigma_{Le}<0.15~\mu$m
   ($\sigma_{Le}>0.15~\mu$m).  }
\label{phase_diag}
\end{figure}

We will now present a
simple mechanical model that can explain all the experimental findings,
quantitatively confirming the existence of a fluctuation increase at a characteristic value of the force
$F_{char}$ and deriving an original relation between $F_{char}$ and
DNA nanomechanical constants (bending constant $B$ and the binding
energy between the DNA bases (denaturation energy)).
The model is an
extension of the classical theory of plectoneme formation
\cite{strick2003} considering the
denaturation energy.
 Indeed, when
 twisted by the magnetic bead, the DNA molecule relaxes the applied work in three different ways:
by twisting (twisting energy: $E_{twist}$), by forming plectonemes (plectonemic energy: $E_{plect}$) and by partially denaturating (denaturation energy: $E_{den}$).
 Accordingly \cite{strick2003}:

\begin{equation}
\label{eq Etwist}
{ E_{twist} = \frac{1}{2} \frac{C}{L_0} (2 \pi \cdot n_e)^2}\\
\end{equation}

\begin{equation}
\label{eq Eplect}
{ E_{plect} = (2  \pi  R \cdot n_p) \cdot (\frac{B}{2 R^2}+F)}\\
\end{equation}

while we have evaluated $E_{den}$ as:

\begin{equation}
\label{eq Eden}
{ E_{den} = \alpha \cdot n_d}\\
\end{equation}

where $L_0$ is the contour length of the DNA, $n_e$ is the number of
turns that store energy in the twist, $n_p$
is the number of plectonemes of radius $R$, $n_d$ is the number of
turns relaxed by partial DNA denaturation, C is the twisting constant, and  $\alpha$ is a phenomenological constant that will be discussed later.
We impose the additional topological relation:  $n_e=n_t-n_p-n_d$.
 Furthermore,
we assume that the DNA
extension variations are mainly due to the plectoneme formation:
$L_e=L_0 - 2 \pi R \cdot n_p$. For this assumption, we disregard for the moment the Worm-Like-Chain (WLC) dependence
of DNA extension on the applied force and we suppose that the denaturation
 phenomenon does not induce a significant DNA length variation.

We can estimate the characteristic force $F_{char}$ by using the
following back-of-the-envelope calculation. The characteristic
force is obtained by equating the plectonemic energy and the
denaturation energy which is necessary for relaxing to the same torsion angle
(i.e. $\Delta n_p$=$\Delta n_d$). The value
 $R =\sqrt{B/2F}$ of the plectonemic radius \cite{strick2003} is kept constant and obtained from the equilibrium of the twisting energy and the plectonemic energy, resulting in the following relations:
\begin{equation}
\label{eq}
{ (2 \pi R) \cdot (\frac{B}{2 R^2}+F) = \alpha }
\end{equation}
obtaining for the characteristic force:
\begin{equation}
\label{eq  Fchar}
{ F_{char} = \frac{\alpha^2}{8 B \pi^2} }
\end{equation}
The parameter $\alpha$  is proportional to the binding energy between the bases of the two DNA
single strands.
Considering the experimental value $F_{char} \approx$~0.78~pN and a DNA persistence length of about 50~nm, the resulting value of $\alpha$ is on the order of $10^{-19}$~J, which is consistent with the reported average value of the free energy per DNA base pair, $\Delta G \approx$~2.0~kcal/mole \cite{Breslauer1986,Santalucia1998,Huguet2010}.
In a first approximation we consider the denaturation energy, described by the
average single parameter $\alpha$, to be independent on the specific DNA
sequence. The resulting value of  $\alpha$
corresponds to approximately 8 denaturated bases to relax one turn. This number of base pairs is consistent with the pitch of the double helix.
More rigorously, fixing the external force $F$ and the total
imposed torsion $n_t$, we can study
the total energy landscape $E_{tot}$ as a function of the parameters
$n_p$, $n_d$, and $R$.
By minimizing the expression of the total energy $E_{tot}$ in the ($n_t$,$F$) plane,
we derive how DNA behaves under imposed torsion and force.
The resulting calculations are presented in Fig.~\ref{phase_diag}A
where the different colored regions represent different DNA behaviors
and the boundary lines correspond to transition lines.
In the white central region, the DNA reaches the maximum extension
compatible with the applied force according to the WLC model \cite{strick2003}.
Entering the lateral blue regions, plectonemic formation starts and DNA extension is consequently reduced. The green regions correspond to achieving zero DNA extension.
The novelty of the model is its capability for predicting the
denaturation regions shown in the red (denaturation bubble) and yellow (coexistence regions between plectonemes and bubbles) rectangles.
The boundary of each zone corresponds to a transition between different structural phases. We concentrate our attention on the transition to the denaturation state.
It is possible to enter the denaturation zone (red and yellow) by crossing three different lines, but only the horizontal line delimiting the red zone is characterized by a step in the  DNA extension. Indeed, below this line the DNA has a reduced length due to the presence of several (depending on the turn position) plectonemes and above this line the DNA is fully extended. Accordingly, the fluctuation between the structures configuration is also accompanied by a oscillation in the DNA extension. This feature is unique to this transition; none of the other phase structure boundaries reported in Fig.~\ref{phase_diag}A introduce any discontinuity in $L_e$.

Furthermore, because $L_e$ depends on $n_p$, calculating $n_p$
using  the proposed expression for the
energy $E_{tot}$ and a Boltzmann distribution allows for the prediction of the average value
of the DNA extension and its fluctuation $\sigma_{Le}$.
The  calculated values of $\langle L_e\rangle$, corrected for the WLC model, are presented as a function of
$n_t$ by the continuous lines in Fig.~\ref{chapeau}, showing a good
agreement with the experimental data.
The results are obtained assuming the following values:
 $B$~=~1.29$\cdot$$10^{-11}$~eV$\cdot$cm, $C$~=~3.62$\cdot$$10^{-11}$~eV$\cdot$cm, $L_0$~=~1.88~$\mu$m and
 $\alpha$~=~0.70~eV.
As shown in Fig.~\ref{chapeau}, the model accurately
describes the classical plectonemic behavior observed for $n_t$~$>$~0, which was already well described in the past \cite{strick2003}, as
well as the transition between the plectonemic and denaturation regime for
$n_t$~$<$~0.

As expected, the region characterized by non-negligible calculated DNA
extension fluctuations $\sigma_{Le}$
is located
in a specific area of the plane ($n_t$,$F$):  at the horizontal boundary of the red zone.
This theoretically predicted region is shown as a contour plot in Fig.~\ref{phase_diag}B (enlargement of a region of Fig.~\ref{phase_diag}A).
In Fig.~\ref{phase_diag}B we also present, as crosses or circles, respectively, the measured standard
deviations having negligible ($\sigma_{Le}$~$<$~0.15 $\mu$m) or significant
($\sigma_{Le}$~$>$~0.15 $\mu$m) values.
From Fig.~\ref{phase_diag}B, we can
appreciate the agreement between our model and the experimental results: the regions of largest fluctuation are observed where
 the model predicts a marginal stability \cite{brogioli2010} of $L_e$, i.e., for
low $n_t$ value ($n_t$~$\leqslant$~-30) and near the characteristic
value of the force $F_{char}$~$\approx$~0.78~pN.

Moreover, in Fig.~\ref{phase_diag}A we sketch vertical and horizontal thick black segments corresponding to the lines explored by the experiments presented in Fig.~\ref{versus_F} (fixed $n_t$) and Fig.~\ref{versus_nt} (fixed F). As predicted by the model, the asymmetry in $L_e$ indicates entrance into the red zone and large fluctuations appear when crossing the red zone horizontal boundary   and when the experiments are performed in its proximity.

In conclusion, we have characterized the denaturation transition caused by external applied force and torsion, and we have introduced a nanomechanical model able to link the measured force at which the denaturation occurs to a parameter $\alpha$ related to the double strand stability.
 Because DNA denaturation and melting is at the origin of several
 biological problems ranging from DNA replication and transcription
 to the detection of transcription initiation points, the ability demonstrated here for a direct, quantitative single molecule
 measurement of the characteristic force and denaturation energy opens the way for several studies
  utilizing more sophisticated and biologically realistic
 situations
 and in the presence of DNA binding molecules or
 proteins \cite{vladescu2005}.

\bibliography{DNA_MT}
\end{document}